\RequirePackage{fix-cm}
\documentclass[twocolumn]{svjour3}          % twocolumn
\smartqed  % flush right qed marks, e.g. at end of proof
\usepackage{floatrow}
\usepackage{amsmath}
\usepackage{graphicx,adjustbox}
\usepackage{subfigure} 
\usepackage{tikz}
\usepackage{gensymb}
\usepackage{amssymb}

\usetikzlibrary{fadings,shapes.geometric,calc,shapes.misc,decorations.text,arrows,positioning,shadows}

\newcommand{\lp}{\vec{l}}				%% L_p	
\newcommand{\pone}{\mathbb{P}}		%% P
\newcommand{\ptwo}{\mathbb{P}^2}		%% P2
\newcommand{\pthree}{\mathbb{P}^3}		%% P3
\newcommand{\rtwo}{\mathbb{R}^2}		%% R2
\newcommand{\rthree}{\mathbb{R}^3}	
\newcommand{\rfourfour}{\mathbb{R}^{4\times4}}		%% R4
\newcommand{\Pvone}{\vec{P_\text{1}}}	%% P1
\newcommand{\Pvonea}{\vec{P}_\text{1,align}}
\newcommand{\Pvoneaa}{\vec{p}^1_\text{1,align}}
\newcommand{\Pvoneab}{\vec{p}^2_\text{1,align}}
\newcommand{\Pvoneac}{\vec{p}^3_\text{1,align}}	%% P1algn
\newcommand{\Pvtwo}{\vec{P_\text{2}}}

\journalname{myjourna}

\definecolor{FAUblue1}{rgb} {0.00000, 0.21961, 0.39608}
\definecolor{FAUblue2}{rgb} {0.56471, 0.65490, 0.77647}
\definecolor{FAUblue3}{rgb} {0.86667, 0.89804, 0.94118}
\begin{document}

\title{Viewpoint Planning for Quantitative Coronary Angiography%\thanks{Grants or other notes
}


\author{Alexander Preuhs %
	\and Martin Berger%
	\and Sebastian Bauer%
	\and Thomas Redel%
	\and Mathias Unberath
	\and Stephan Achenbach%
	\and Andreas Maier%  %etc.
}


\institute{Alexander Preuhs \and Mathias Unberath \and Andreas Maier\at
              Pattern Recognition Lab,  Friedrich-Alexander Universit\"at Erlangen-N\"urnberg,   Germany\\
              \email{alexander.preuhs@fau.de}           %  \\
           \and
           Martin Berger \and Sebastian Bauer \and Thomas Redel \at
              Siemens Healthcare GmbH, Forchheim, Germany
            \and
            Stephan Achenbach \at
            Department of Cardiology, Universit\"atsklinikum Erlangen, Germany
}

\date{Received: 27 January 2018 / Accepted: 3 April 2018 / Published online: 1 June 2018}

\maketitle

\begin{abstract}%%
	~\\
	\textbf{Purpose:} In coronary angiography the condition of myocardial blood supply is assessed by analyzing \mbox{2-D} X-ray projections of contrasted coronary arteries. This is done using a flexible C-arm system. Due to the X-ray immanent dimensionality reduction projecting the \mbox{3-D} scene onto a \mbox{2-D} image, the viewpoint is critical to guarantee an appropriate view onto the affected artery and, thus, enable reliable diagnosis. In this work we introduce an algorithm computing optimal viewpoints for the assessment of coronary arteries without the need for \mbox{3-D} models. \\
	\textbf{Methods:} 	We introduce the concept of optimal viewpoint planning solely based on a single angiographic X-ray image. The subsequent viewpoint is computed such that it is rotated precisely  around a vessel, while minimizing foreshortening.\\
	\textbf{Results:} Our algorithm reduces foreshortening substantially compared to the input view and completely eliminates it for $90\degree$ rotations. Rotations around  iso-centered foreshortening-free vessels passing the isocenter are exact. The precision, however, decreases when the vessel is off-centered or foreshortened. We evaluate worst case boundaries, providing insight in the maximal inaccuracies to be expected. This can be utilized to design viewpoints guaranteeing desired requirements, e.g. a true rotation around the vessel of at minimum $30\degree$. In addition a phantom study is performed investigating the impact of input views to 3-D quantitative coronary angiography (QCA).\\ 
	\textbf{Conclusion:} We introduce an algorithm for optimal viewpoint planning from a single angiographic X-ray image. The quality of the second viewpoint~---~i.e. vessel-foreshortening and true rotation around vessel~---~depends on the first viewpoint selected by the physician, however, our computed viewpoint is guaranteed to reduce the initial foreshortening.
	Our novel approach uses fluoroscopy images only and, thus, seamlessly integrates with the current clinical workflow for coronary assessment. In addition it can be implemented in the QCA workflow without increasing user-interaction, making vessel-shape reconstruction more stable by standardizing viewpoints. 
	
\end{abstract}

\section{Introduction}
\label{introduction}
Cardiovascular diseases constitute a growing global health problem. In the USA 31.1 \%  of all death in 2011 are categorized as related to cardiovascular disease, whereby half of the incidents are accountable to Coronary Heart Disease (CHD) \cite{Mozaffarian2015}. The diagnosis of CHD is commonly performed by evaluating the vessel shape  of coronary arteries after contrast agent injection. 
While many imaging modalities allow for the assessment of vessel shape, such as CT-Angiography or ultrasound, catheter-based X-ray angiography using interventional C-arm cone-beam systems is still considered the work-horse modality, as it allows for diagnosis and treatment in a single session and procedure. 
However, the \mbox{3-D} vessel structure is to be assessed on \mbox{2-D} projection images. This may lead to false interpretations due to projective simplifications. Thus, the proper selection of viewpoints is important. Currently, viewpoints are determined as standard angulations followed by iterations of manual adjustments, causing unnecessary dose and contrast injection. 

An approach to improve the \mbox{2-D-based} assessment is described by \mbox{3-D} quantitative coronary angiography (QCA), where two projection images~---~acquired within the same heart phase~---~are used to perform a symbolic \mbox{3-D} reconstruction of a vessel segment. The reconstruction is based on finding corresponding points on the centerlines of the vessel and using a vessel segmentation to approximate the lumen while enforcing consistency \cite{delaere1991,Garcia2009,Liu1992,Pellot1994}. 

However, the quality of the \mbox{3-D} QCA is dependent on the proper selection of the two input projections. The X-ray projection inherently performs a dimensionality reduction, whereas most information is lost in the depth direction. To preserve most of the relevant information needed for the assessment of the vessel, the projection direction must be selected accordingly. Thus, the second viewpoint should be rotated around the vessel, which should be parallel to the detector in both viewpoints \cite{Sato1998}.

The selection of good viewpoints is not trivial and often a source of inaccuracy as the physician needs to project the vessel without foreshortening and rotated around the vessel that is to be assessed. Green et al. compared the amount of foreshortening in physician selected viewpoints with the viewpoints generated from available \mbox{3-D} coronary trees \cite{Green2005}. Summarizing the average vessel foreshortening was greater than 20\% in 18\% of the cases and below 10\% in  64\% of the cases. The resulting cases showed a foreshortening between 11\% and 19\%. A second source of error is the amount of rotation around the vessel. Commercially available systems measure the position of the C-arm using a latitude-longitude spacing, where the latitude and longitude are expressed as CRAN/CAUD and RAO/LAO angle (cf. Fig. \ref{fig:carmgeometry}). The drawback of this equal angle systems is that the great circle distance does not necessarily equal the amount of rotation defined by RAO/LAO. For example, consider a C-arm that is angulated by $45\degree$ in the CRAN direction, an additional RAO rotation by $30\degree$ will result in an overall rotation of only $21.09\degree$ measured between the principal rays of the two views. Generally speaking this effect vanishes for CRAN/CAUD $= 0\degree$ and is getting more intense for higher initial CRAN/CAUD angulations.

Viewpoint planning for coronary angiography was discussed in literature to obtain optimal viewpoints \cite{JamesChen2000,Tu2010} or create view-maps which can be used as a heuristic look-up-table to identify angulations that are probably best suitable for a particular vessel segment \cite{Garcia2009,Wink2002}. Fallavollita et al. presented the highly related concept of 'desired view' in angiographic interventions \cite{fallavollita2014}. Based on a pre-operative CTA the physician only communicates the desired view whereas the C-arm position is automatically provided by the system. This concept reduces dose, as no unnecessary X-ray projections are captured during the manual positioning process. This concept was later applied to aortic interventions \cite{virga2015}.

All previous methods rely on a \mbox{3-D} reconstruction of the vessel. However, the \mbox{3-D} reconstruction itself requires proper selection of input views or additional pre-interventional imaging \cite{Unberath2017}. 
A highly related method was proposed by Chrisriaens et al. \cite{Chrisriaens2001}. The method does not require a \mbox{3-D} reconstruction, instead they determine optimal viewpoints for the determination of QCA parameters based on two projection images acquired with a Bi-plane system. The target vessel from which the QCA parameters should be calculated is selected on both projections, then the orientation of the vessel is calculated and the optimal viewpoint is determined. 
To the best of our knowledge, no method addresses the determination of optimal viewpoints based on a single \mbox{2-D} projection. 

We introduce a viewpoint planning system that only uses \mbox{2-D} information from a single angiographic projection. The main purpose is to find a second optimal view that can be used~---~together with the initial projection~---~for \mbox{3-D} QCA. The second view is rotated around the vessel by a given angle while minimizing  vessel foreshortening. However, as depth information is missing, the accuracy is dependent on the positioning of the vessel within the first view, i.e. amount of foreshortening and offset from the isocenter in view direction. To evaluate the clinical applicability of the \mbox{2-D} based viewpoint planning an accuracy evaluation is performed, showing the restrictions of the presented algorithm. The dependency of the input views on the quality of \mbox{3-D} QCA is investigated with a phantom study in an interventional environment.

\section{Materials and Methods}
\label{methods}
At the beginning of this section, we introduce the conventions used for describing the C-arm geometry. This includes the description of the C-arm system itself and the mathematics describing the acquisition of X-ray images. Thereafter, the  proposed viewpoint planning algorithm is described. 

\subsection{Geometry of a C-arm System}

During X-ray acquisitions using a C-arm system, an X-ray source radially emits photons in a cone-like profile which are then registered at a detector. The geometry describing the relation between a point  $\vec{x} \in \rthree$ within that cone and its corresponding projection on the detector $\vec{u} \in \rtwo$  can be described mathematically by a perspective transformation. 

A perspective transform can be expressed elegantly in terms of matrix multiplication in the projective domain $\pone$. For world points $(x,y,z)^\top \in \rthree$ we will use the homogeneous representation $(x,y,z,w)^\top \in \pthree$ where $w$ describes the homogeneous component being 1. Analogously, detector points $(u,v)^\top \in \rtwo$ are described by their homogeneous representation $(u,v,w)^\top \in \ptwo$ with $w = 1$.

In the projective space, we can assign two interpretations for the same object. The vector $(a,b,c,d)^\top \in \pthree$ can either be interpreted as a point or a plane. The Euclidean representation of the plane $(a,b,c,d)^\top \in \pthree$  is defined by all points $x,y,z \in \rthree$ satisfying $ax+by+cz+d =0$. Analogously the vector $(a,b,c)^\top \in \ptwo$ can be interpreted as either a point or a line, where the Euclidean interpretation of the line $(a,b,c)^\top \in \ptwo$ is defined by all points $x,y \in \rtwo$ satisfying $ax+by+c = 0$. 

A special case is the representation of a line in $\pthree$. Opposite to the previous geometric quantities there is no direct description but we can construct the line as the connection of two points or the intersection of two planes.   
We call the connection of two geometric objects ``join" and the intersection ``meet". 
For $\ptwo$ we define the operations on the vectors $\vec{a}, \vec{b} \in \ptwo$ as
\begin{equation}
\label{eq:antisym}
\text{meet}(\vec{a},\vec{b}) = \text{join}(\vec{a},\vec{b}) = [ \vec{a} ]_\times \vec{b}   = \begin{pmatrix}
0 && -a_z && a_y \\
a_z&& 0 && -a_x \\
-a_y && a_x&&0
\end{pmatrix} 
\vec{b}
\enspace.
\end{equation}
In the $\pthree$, the meet of two points or the join of two planes is more challenging. An intuitive derivation can be found in \cite{Blinn1977}, we will only state the result of this derivation. Both operations result in a line $\vec{L}$  defined by 6 parameters $(p,q,r,s,t,u)^\top$ that are  commonly referred to as Pl\"ucker coefficients. They can be calculated from $\vec{a}, \vec{b} \in \pthree$ by
\begin{equation}
\label{eq:join_meet_p3}
\text{meet(\vec{a},\vec{b})} = \text{join(\vec{a},\vec{b})} =
\vec{L} = 
\begin{pmatrix}
p \\ q \\ r\\ s\\t \\u 
\end{pmatrix} 
= 
\begin{pmatrix}
a_z b_w - a_w b_z\\ a_y b_w - a_w b_y \\ a_y b_z - a_z b_y\\ a_x b_w - a_w b_x\\ a_x b_z - a_z b_x \\ a_x b_y - a_y b_x 
\end{pmatrix} \enspace.
\end{equation}
Again, $\vec{a}, \vec{b}$ can either be two points or two planes. Analogously to $\ptwo$ we can build up anti-symmetric matrices (cf. Eq. \eqref{eq:antisym}) to compute the meet between a line $\vec{L} \in \pthree$ and a plane $\vec{a} \in \pthree$  by
\begin{equation}
\label{eq:meetlineplane}
\text{meet}(\vec{L},\vec{a}) =  \vec{L}^K \, \vec{a} =
\begin{pmatrix}
0 & -u & - t & -s\\u & 0 & -r & -q\\ t & r & 0& -p\\ s & q & p & 0
\end{pmatrix}  \, \vec{a}
\end{equation}
and the join between a line $\vec{L} \in \pthree$ and a point $\vec{b} \in \pthree$ by
\begin{equation}
\label{eq:joinlinepoint}
\text{join}(\vec{L},\vec{b}) = \vec{b}^\top \, \vec{L}^L = \vec{b}^\top \,
\begin{pmatrix}
0 & p & -q & r\\ -p & 0 & s & -t\\ q & -s & 0 & u\\-r & t &-u&0
\end{pmatrix} 
\enspace. 
\end{equation}
Note that Eq. \eqref{eq:meetlineplane} will result in a point and Eq. \eqref{eq:joinlinepoint} defines a plane. We refer to $\vec{L}^K$ as the dual and $\vec{L}^L$ as the primal representation of a line.
The presented operations enable the effortless description of a geometrical scene in the context of a C-arm system, where we want to switch between $\pthree$ and $\ptwo$ corresponding to the world coordinate system and the detector coordinate system.

A well known projective transformation describing the relation between world points and their projection is given by the $3\times4$ projection matrix $\vec{P}$ which originates from the pinhole camera model. The matrix incorporates the whole geometry of the C-arm system.

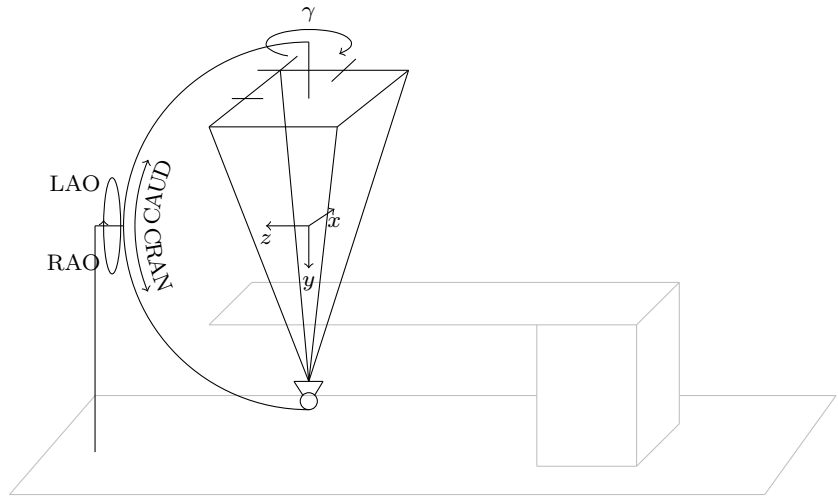
\begin{figure*}
	\floatbox[{\capbeside\thisfloatsetup{capbesideposition={left,top},capbesidewidth=4cm}}]{figure}[\FBwidth]
	{\caption{Schematic drawing of the C-arm geometry. The coordinate axes correspond to the isocenter of the C-arm system. The C-arm system is shown in default position and can be rotated along the drawn rotation axes i.e. right anterior oblique (RAO) and left anterior oblique (LAO) as well as cranial (CRAN) and caudal (CAUD). In addition a detector rotation $\gamma$ can be performed. For coronary interventions, the patient is lying with the head near the isocenter and the feet pointing in the negative $z$ direction.}\label{fig:carmgeometry}}
	{\begin{tikzpicture}[scale=0.375]
		%floor
		\draw[lightgray]  (-11,-8)  edge (-14,-11.5);
		\draw[lightgray]  (-14,-11.5) edge (12.5,-11.5) ;
		\draw[lightgray]  (4.5,-8)  edge (-11,-8) ;
		\draw[lightgray] (9.5,-8) edge (15,-8);
		\draw[lightgray] (15,-8) edge (12.5,-11.5) ;
		
		%Table
		\draw[lightgray]  (-7,-5.5) edge (8,-5.5);
		\draw[lightgray]  (-5.5,-4)  edge (9.5,-4);
		\draw[lightgray]  (9.5,-4) edge (8,-5.5);
		\draw[lightgray]  (-5.5,-4)  edge (-7,-5.5);

		%Table Stand
		\draw[lightgray]   (4.5,-5.5) edge (4.5,-10.5);
		\draw[lightgray]  (8,-10.5) edge (4.5,-10.5);
		\draw[lightgray]  (8,-10.5) edge (8,-5.5);
		\draw[lightgray]  (8,-10.5) edge (9.5,-9);
		\draw[lightgray]  (9.5,-9) edge (9.5,-4);
		%%detector
		\draw   (-4.5,3.5) edge (-7,1.5);
		\draw  (-7,1.5)edge (-2.5,1.5);
		\draw  (-2.5,1.5) edge (0,3.5);
		\draw  (0,3.5) edge  (-4.5,3.5);

		%%detector rays
		\draw  (-7,1.5) edge (-3.5,-7.5);
		\draw  (-2.5,1.5) edge (-3.5,-7.5);
		\draw  (0,3.5) edge (-3.5,-7.5);
		\draw   (-4.5,3.5) edge (-3.5,-7.5);
		
		%Carm
		\draw (-3.5,-8.5) node (v8) {} arc (-90:-180:6.5);
		\draw (-10,-2) node (v9) {} arc (180:90:6.5);
		
		%Carm to Detector
		\draw  (-3.5,4.5) edge (-3.5,2.5);
		%Carm to SOurce
		%\draw  (-3.5,-7.5)edge (-3.5,-8.5);
		
		%Stand to Carm
		\draw  (-10,-2) edge (-11,-2);
		\draw  (-11,-2) edge(-11,-10);
		
		%%gamma
		\draw [->] plot[smooth, tension=.7] coordinates {(-5,4.4) (-4.98,4.5) (-4.9,4.62) (-4.8,4.69) (-4.7,4.74) (-4.6,4.78) (-4.4,4.84) (-4.2,4.88) (-3.9,4.92) (-3.7,4.94) (-3.5,4.94) (-3.3,4.94) (-3.05,4.92) (-2.8,4.89) (-2.6,4.85) (-2.43,4.8) (-2.24,4.72) (-2.09,4.61) (-2.01,4.5) (-2.01,4.39) (-2.06,4.3) (-2.16,4.22) (-2.21,4.19) (-2.31,4.14) (-2.43,4.08)};
		\draw [] plot[smooth, tension=.7] coordinates { (-4.23,4.01) (-4.35,4.03) (-4.44,4.05) (-4.58,4.1) (-4.69,4.14) (-4.8,4.2) (-4.89,4.26) (-4.95,4.31) (-5,4.4)};

		%markierung Detector
		%\node (v64) at (-5.3,3.5) {};
		\draw  (-4.5,3.5) edge (-5.3,3.5);
		%\node at (-4.5,3.5) {};
		\draw (-6.2,2.5) edge (-5.1,2.5);
		\draw (-1.85,3.9)  edge (-2.7,3.1);
		\draw  (-3.9,4) edge (-4.5,3.5);
		
		%RAO LAO
		\draw  (-10.4,-2) ellipse (0.3 and 1.7);
		\draw [->] plot[smooth, tension=.7] coordinates {(-10.7,-1.8)};
		
		%CAUD
		\draw [->,postaction={decorate,decoration={raise=-2.5ex,text along path,text align=center,text={|| CAUD}}}](-9.6,-2) arc (180:157.5:6.1);
		%CRAN
		\draw[->,postaction={decorate,decoration={raise=1ex,text along path,text align=center,text={|| CRAN}}}] (-9.6,-2) arc (-179.9999:-158:6.2);
		%\draw [->](-9.6,-2) arc (180:202.5:6.1);
		
		%Y axis
		\draw [->] (-3.5,-2) -- (-3.5,-3.5)node[below]{$y$} ;
		%Zaxis
		\draw [->] (-3.5,-2) -- (-5,-2)node[below]{$z$};
		%Xaxis
		\draw [->] (-3.5,-2) -- (-2.6,-1.4)node[below]{$x$} ;
		
		\node [left] at (-10.5,-0.5) {LAO};
		\node [left] at (-10.5,-3.3) {RAO};
		\node [above] at  (-3.5,4.94) {$\gamma$};
		
		\draw  (-3.5,-8.2) ellipse (0.3 and 0.3);
		\draw  (-4.02,-7.5) edge (-3,-7.5);
		\draw (-3,-7.5) edge (-3.3,-8);
		\draw (-4.02,-7.5) edge (-3.7,-8);
		\end{tikzpicture}}
\end{figure*}

A schematic drawing of a C-arm system is depicted in Fig. \ref{fig:carmgeometry}.
The main characterization is the movable C-shaped detector-X-ray-source configuration, enabling \mbox{2-D} X-ray projections with a high flexibility. The C-arm can be rotated in CRAN/CAUD and RAO/LAO direction. The rotation can be incorporated into the projection matrix by right multiplication of rotation matrices. If $\vec{P}_0$ corresponds to the C-arm orientation in the initial position (cf. Fig. \ref{fig:carmgeometry}) then a rotated view is described by the angles RAO/LAO$=\alpha$, CRAN/CAUD$=\beta$ and a detector rotation $\gamma$. The rotated view is calculated by
\begin{equation}
\label{eq:carm_rotation}
\vec{P}(\alpha,\beta,\gamma) = \vec{P}_0\, \vec{R}_y(\gamma) \, \vec{R}_x(\beta) \,, \vec{R}_z(\alpha) \enspace,
\end{equation}
where $\vec{R}_{e}(\cdot) \in \rfourfour$ is the homogeneous representation of  a rotation matrix, describing the rotation around a coordinate axis taking the angle as an argument. For the computation of $\alpha$ and $\beta$ from $\vec{P}$ we use the property that the vector $(P_{31}, P_{32},P_{33})^\top$ points in the direction of the principal ray. Thus, by normalizing that vector we obtain $n_x$, $n_y$ and $n_z$ which can be used to calculate the angulation from $\vec{P}$ by
\begin{align}
\label{eq:aplhabeta}
\alpha = \text{atan2}(n_x, -n_y) \, \frac{180\degree}{\pi} && \beta = \text{arcsin}(n_z) \, \frac{180\degree}{\pi}
\end{align}
In the remainder of this manuscript we will mostly talk about projection matrices, however, using Eq. \eqref{eq:aplhabeta} we can easily transform the projection matrix to angles, which could be used to steer a C-arm. Further note that we do not explicitly compute the image rotation $\gamma$, as this is typically not performed in the procedures and is always kept at $\gamma = 0\degree$ or $\gamma = 90\degree$.

%%%%%%%%%%%%%%%%%%%%%%%%%%%%%%%%%%%%%%%%%%%%%%%%%%%%%%%%%%%%%%%%%%%%%%%%%%%%%%%%%%%%%%%%%%%%%%%%%%
%%%%%%%%%%%%%%%%%%%%%%%%%%%%%%%%%%%%%%%%%%%%%%%%%%%%%%%%%%%%%%%%%%%%%%%%%%%%%%%%%%%%%%%%%%%%%%%%%%
%%%%%%%%%%%%%%%%%%%%%%%%%%%%%%%%%%%%%%%%%%%%%%%%%%%%%%%%%%%%%%%%%%%%%%%%%%%%%%%%%%%%%%%%%%%%%%%%%%
%%%%%%%%%%%%%%%%%%%%%%%%%%%%%%%%%%%%%%%%%%%%%%%%%%%%%%%%%%%%%%%%%%%%%%%%%%%%%%%%%%%%%%%%%%%%%%%%%%
%%%%%%%%%%%%%%%%%%%%%%%%%%%%%%%%%%%%%%%%%%%%%%%%%%%%%%%%%%%%%%%%%%%%%%%%%%%%%%%%%%%%%%%%%%%%%%%%%%
%%%%%%%%%%%%%%%%Start of Method

\subsection{2-D-Based Viewpoint Planning}
We introduce a method for viewpoint planning based on a single \mbox{2-D} angiographic X-ray image. The second viewpoint is expected to be rotated around the vessel by a physician determined angle and with the vessel being projected without foreshortening.  

In the simplest case, the target vessel is in the C-arm isocenter and not foreshortened in the initial X-ray projection. While this assumption will almost never be satisfied in clinical practice, it is a good starting point to grasp on the general idea. First, we estimate the rotation axis of our transformation. In this very simple case, assuming a rotation axis that is simply the backprojection of the vessel to the isocenter, will produce exact results as we know that the vessel of interest is a) in the isocenter and b) not foreshortened. Thus, we can use this axis and rotate the gantry around it.

However, in clinical practice the target vessel is not necessarily in the isocenter, nor is it projected without foreshortening. This makes an exact determination of the rotation axis infeasible as depth cannot be recovered from a single image, but we can make some adjustments to perform substantially better than by just assuming a centered, foreshortening-free vessel. 
Fig. \ref{fig:fc_vp} depicts the steps that build up our workflow which are explained in the following subsections. 

\subsubsection{Principal Ray Alignment}
\label{sec:alignment}
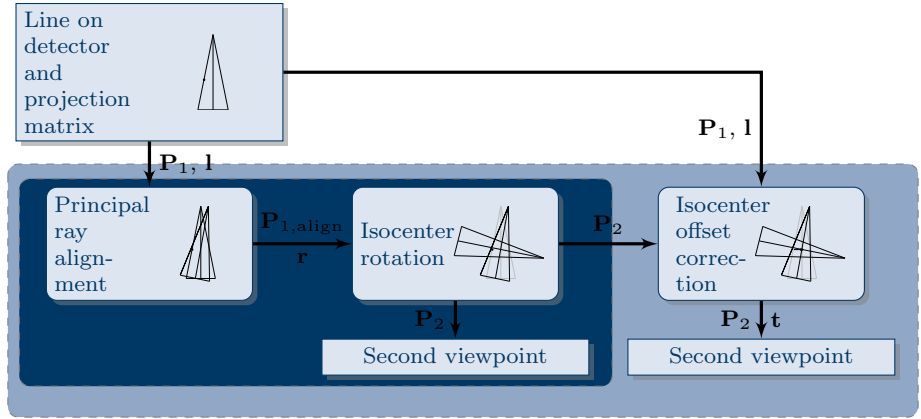
\begin{figure*}
	\centering
	\usetikzlibrary{fadings,shapes.geometric,calc,shapes.misc,decorations.text,arrows,positioning,shadows}

%	\tikzstyle{blockmath}=[rectangle, draw, fill=blue!20, 
%	text width=24mm, text badly centered, rounded corners,
%	minimum height=12mm,drop shadow]
	
	\tikzstyle{block}=[rectangle, draw, fill=FAUblue3, 
	text width=25mm, text badly centered, rounded corners,
	minimum height=15mm,drop shadow]
	
	\tikzstyle{block2}=[rectangle, draw, fill=FAUblue3, 
	text width=33mm, text badly centered, 
	minimum height=1mm,drop shadow]
	\tikzstyle{line}=[draw, very thick, color=black, -latex']
	\pgfdeclarelayer{background}
	\pgfdeclarelayer{foreground}
	\pgfsetlayers{background,main,foreground}

	\tikzstyle{blockVP}=[rectangle, draw, fill=FAUblue3, 
	text width=50mm, text badly centered, rounded corners,
	minimum height=20mm,drop shadow]
	\tikzstyle{add}=[circle, draw, fill=blue!20, 
	text width=8mm, text badly centered, rounded corners,
	minimum height=8mm,drop shadow]
	
	\begin{tikzpicture}
%	\definecolor{FAUblue1}{rgb} {0.00000, 0.21961, 0.39608}
%	\definecolor{FAUblue2}{rgb} {0.56471, 0.65490, 0.77647}
%	\definecolor{FAUblue3}{rgb} {0.86667, 0.89804, 0.94118}
	\color{FAUblue1}
	\node [block2] (start)
	{
		\noindent\begin{minipage}{0.5\textwidth}\raggedright
		Line on detector and projection matrix
		\end{minipage}%
		\begin{minipage}{0.5\textwidth}\centering
		\includegraphics[height = 10 mm]{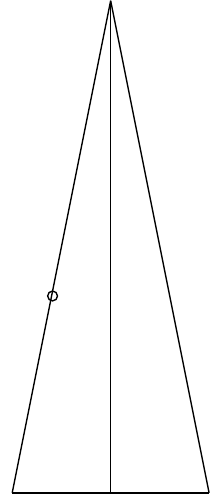} 
		\end{minipage}
	};
	\node [block, below=6mm of start] (algn_view)	{
		\noindent\begin{minipage}{0.5\textwidth}\raggedright
		Principal ray alignment
		\end{minipage}%
		\begin{minipage}{0.5\textwidth}\centering
		\includegraphics[height = 10 mm]{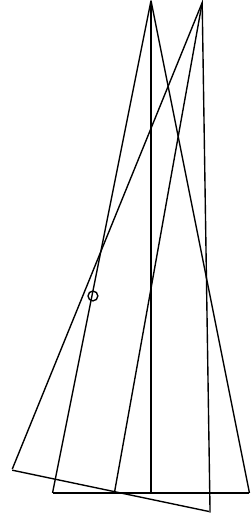} 
		\end{minipage}
	};
	\node [block, right=13mm  of algn_view] (rotxi)
	{
		\noindent\begin{minipage}{0.45\textwidth}\raggedright
		Isocenter rotation
		\end{minipage}%
		\begin{minipage}{0.55\textwidth}\centering
		\includegraphics[height = 10 mm]{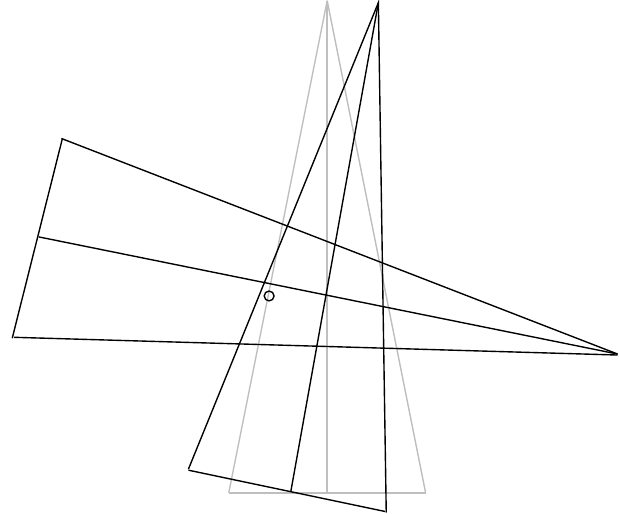} 
		\end{minipage}
	};
	\node [block, right=13mm  of rotxi] (tbl_mve)
	{
		\noindent\begin{minipage}{0.4\textwidth}\raggedright
		Isocenter offset correction
		\end{minipage}%
		\begin{minipage}{0.5\textwidth}\centering
		\includegraphics[height = 10 mm]{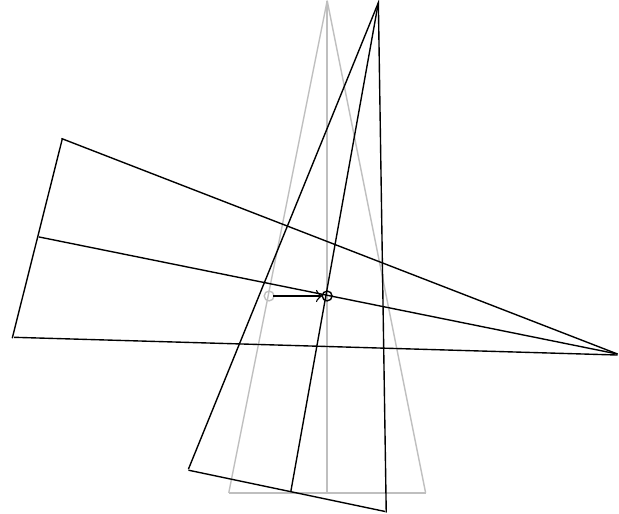} 
		\end{minipage}
	};
	\node [block2, below=5mm of tbl_mve] (end){Second viewpoint};
	\node [block2, below=5mm of rotxi] (endnotbl){Second viewpoint};
	
	\path [line] (start) --node[right]{$\Pvone$, $\lp$} (algn_view);
	\path [line] (algn_view) -- node[above]{$\Pvonea$ }node[below]{$\vec{r}$ } (rotxi);
	\path [line] (rotxi) -- node[above]{$\Pvtwo$}node[below]{} (tbl_mve);
	\path [line] (tbl_mve) -- node[left]{$\Pvtwo$} node[right]{$\vec{t}$}(end);
	\path [line] (rotxi) -- node[left]{$\Pvtwo$} (endnotbl);
	
	\path [line] (start) -| node[near end,left]{$\Pvone$, $\lp$} (tbl_mve);
	
	\begin{pgfonlayer}{background}
	\path (algn_view.west |- tbl_mve.north)+(-0.5,0.3) node (a) {};
	\path (end.south -| end.east)+(+0.3,-0.55) node (b) {};
	\path[fill=FAUblue2,rounded corners, draw=black!50, dashed]
	(a) rectangle (b);
	\end{pgfonlayer}
	
	\begin{pgfonlayer}{background}
	\path (algn_view.west |- tbl_mve.north)+(-0.35,0.105) node (a) {};
	\path (endnotbl.south -| endnotbl.east)+(+0.3,-0.14) node (b) {};
	\path[fill=FAUblue1,rounded corners, draw=black!50, dashed]
	(a) rectangle (b);
	\end{pgfonlayer}
%	\node (vp_calc) [below=11mm of rotxi] {Section \ref{sec:rotation}};
%	\node (vp_calc) [below right=11mm and -63mm of rotxi] {Section \ref{sec:alignment}};
%	\node (vp_calc) [below right=11mm and 17mm of rotxi] {Section \ref{sec:isocenter}};
	\end{tikzpicture}
	\caption{Flowchart showing the proposed viewpoint planning algorithm. The input is a line defined on the detector. A principal ray alignment is performed, creating a virtual intermediate view. The isocenter rotation is then applied to the virtual intermediate view. The algorithm is finished after the calculation of an additional translation compensating for an isocenter offset. If a translation cannot be performed, e.g. due to limitations of the system, the isocenter offset correction can be skipped.}
	\label{fig:fc_vp}
\end{figure*}
The starting point of the algorithm is a \mbox{2-D} projection image of contrasted vessels, and the corresponding projection matrix $\vec{P}_1$. The physician selects the target vessels by two clicks $\vec{c_1}, \vec{c_2} \in \ptwo$ defining a line $\vec{l} = \text{join}(\vec{c_1},\vec{c_2})$. Vessels within the isocenter that are not foreshortened are already well aligned, making the principal ray alignment dispensable. However, in most cases, this will not be the case. 

Assume that the vessel is off-centered e.g. $5\,$mm parallel to the detector. In this case, due to the X-ray cone, we already observe the vessel from an orientation rotated compared to the principal ray. This rotation is what we seek to compensate with a principal ray alignment.
Therefore, we first calculate the backprojection-plane of the target vessel $\vec{e}_L$ and compare its normal to that of a second plane $\vec{e}_R$ which is constructed by the backprojection of $\vec{c_1}$ and $\vec{c_2}$ as well as the source position. With $\vec{p}_1^i$ denoting the $i-$th row of the projection matrix $\vec{P_1}$, we can interpret each $\vec{p}_1^i$ as a plane, each passing the source position. Therefore, we can compute the source position $\vec{s} \in \pthree$ simply by $\vec{s} = \text{meet}(\text{meet}(\vec{p}_1^1, \vec{p}_1^2),\vec{p}_1^3)$. The two planes are then calculated by
\begin{align}
\vec{e}_L = \vec{P}_1^{+\top} \vec{l} && \vec{e}_R = \text{join}(\text{join}(\vec{P}_1^{+\top}\vec{c_1}, \vec{P}_1^{+\top}\vec{c_2}), \vec{s}) \enspace ,
\end{align}
note that $\vec{e}_L$ and $\vec{e}_R$ are equal, if the vessel is in the isocenter, but will differ when the vessel is translated. The angle $\alpha$ between the planes $\vec{e}_L$ and $\vec{e}_R$ can simply be calculated from the angle between their normals, e.g. by exploiting the definition of the scalar product. 
In the next step we use the property, that we can right-multiply rotation matrices $\vec{R}\in \rfourfour$ to $\Pvone$. The resulting projection matrix corresponds to a virtual intermediate view rotated by $\vec{R}$. Therefore, we transform the rotation around an axis $\vec{r}$ by $\alpha$ to a rotation matrix $\vec{R}_{\vec{r},\alpha}$ using a homogeneous version of the Rodriguez formula. The axis $\vec{r}$ is the orientation of the backprojected vessel, which will be discussed in more detail in the following subsection (cf. Eq. \eqref{eq:defR}). This is then right-multiplied to $\Pvone$ in order to obtain the intermediate view
\begin{equation}
\Pvonea = \Pvone \vec{R}_{\vec{r},\alpha} \enspace.
\end{equation}
As depicted in Fig. \ref{fig:fc_vp}, the aligned intermediate view $\Pvonea$ will be propagated to the "isocenter rotation" module.

\subsubsection{Isocenter Rotation}
\label{sec:rotation}
The goal is now to rotate around the vessel segment as exact as possible using the intermediate view. The rotation axis is the backprojection of the line $\vec{l}$ defined on the detector. 
Note that the backprojection operation basically solves $\vec{P} \vec{x} = \vec{y}$ with $\vec{x}$ denoting world points and $\vec{y}$ detector points. Since this equation is not uniquely solvable we use the pseudo-inverse $\vec{P}^+$ to calculate $\vec{x}$. If the principle ray of the system intersects the coordinate origin, then the space of solutions to $\vec{P}^+ \vec{y}$ corresponds to a plane parallel to the detector. This justifies our assumption, that we can use the backprojection of the line $\vec{l}$ as rotation axis. We can interpret $[\vec{l}]_\times$ as a tensor which is therefore transformed as such. With the backprojection $\vec{P}_1^+$ being the desired transformation, the backprojection of the line $\vec{l}$ is computed by
\begin{equation}
\vec{L}^K \, = \, \vec{P}_1^+\, [\vec{l}]_\times\,  \vec{P}_1^{+\top} \enspace.
\end{equation}
The result $\vec{L}^K$ is the dual representation of the line $\vec{L}$  that contains the Pl\"ucker coordinates in its respective entries (cf. Eq. \eqref{eq:meetlineplane}), thus we can simply extract $\vec{L}$ from $\vec{L}^K$. The direction of $\vec{L}$ can be found by the intersection with the plane at infinity $\pi^\infty = (0,0,0,1)^\top$, which will be a point  $\vec{r}\in \pthree$ at  infinity
\begin{equation}
\vec{r} = 	\text{meet}(\vec{L},\pi^\infty) \enspace.
\label{eq:defR}
\end{equation} 
The first three components of $\vec{r}$ denote the direction of $\vec{L}$, which is the rotation axis we aim to compute. 
Finally, we create the second viewpoint $\vec{P}_2$ by rotating around $\vec{r}$ by $\xi$, again using Rodriguez formula
\begin{equation}
\vec{P}_{2} = \Pvonea \, 	\vec{R}_{\vec{r},\xi} 
\enspace .
\label{eq:rodrig}
\end{equation}

\begin{figure*}
	\floatbox[{\capbeside\thisfloatsetup{capbesideposition={left,top},capbesidewidth=4cm}}]{figure}[\FBwidth]
	{\caption{Heatmap showing the maximal foreshortening in the second view for vessels having an initial foreshortening of $\nu = 30^\circ$.}	\label{fig:fs_accuracy}}
	{
		\centering
		\resizebox{1.18\linewidth}{!}{
			\input{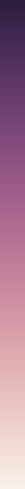}}}
\end{figure*}
\subsubsection{Isocenter Offset Correction}
\label{sec:isocenter}
The input of this algorithm is $\Pvtwo$ described in the previous sections.
To compute $\Pvtwo$ we have rotated around an isocenter, that did not correspond to the center of the vessel. 
In fact, we cannot know the true center of rotation, i.e. the center of the vessel segment, as \mbox{3-D} reconstruction of the vessel is impossible from the single frame only. Yet, we can minimize the difference between the true isocenter and the isocenter of rotation. Doing so results in a translation $\vec{t}$ that can be realized by either a table movement or a C-arm translation. The goal of the offset correction, is that the central ray of $\Pvonea$ is coincident with the backprojection-plane of the vessel  $\vec{e}_L$. Therefore, we calculate the distance between $\vec{e}_L$ and the central ray of $\Pvonea$. With the plane and line being parallel by construction, this distance is equal to the distance between $\vec{e}_L$ and the source position of $\Pvonea$. We denote the source position of $\Pvonea$ by $\vec{s}_a = \text{meet}$$(\text{meet}(\Pvoneaa,\Pvoneab),\Pvoneac)$ and calculate the distance $d$ by
\begin{equation}
\label{eq:plane_distance}
d = \frac{\vec{e}_L^T}{\sqrt{\vec{e}_{L_1}^2 + \vec{e}_{L_2}^2 + \vec{e}_{L_3}^2}}	  \, 
\frac{\vec{s}_a}{\sqrt{\vec{s}_{a_1}^2 + \vec{s}_{a_2}^2 +\vec{s}_{a_1}^3}} \, \enspace.
\end{equation}  
Note that Eq. \eqref{eq:plane_distance} is simply the distance between a point and a plane. The translation $\vec{t}$ is then in the normal direction of $\vec{e}_L$ scaled by $-d$.

\section{Evaluation and Results}
\begin{figure*}
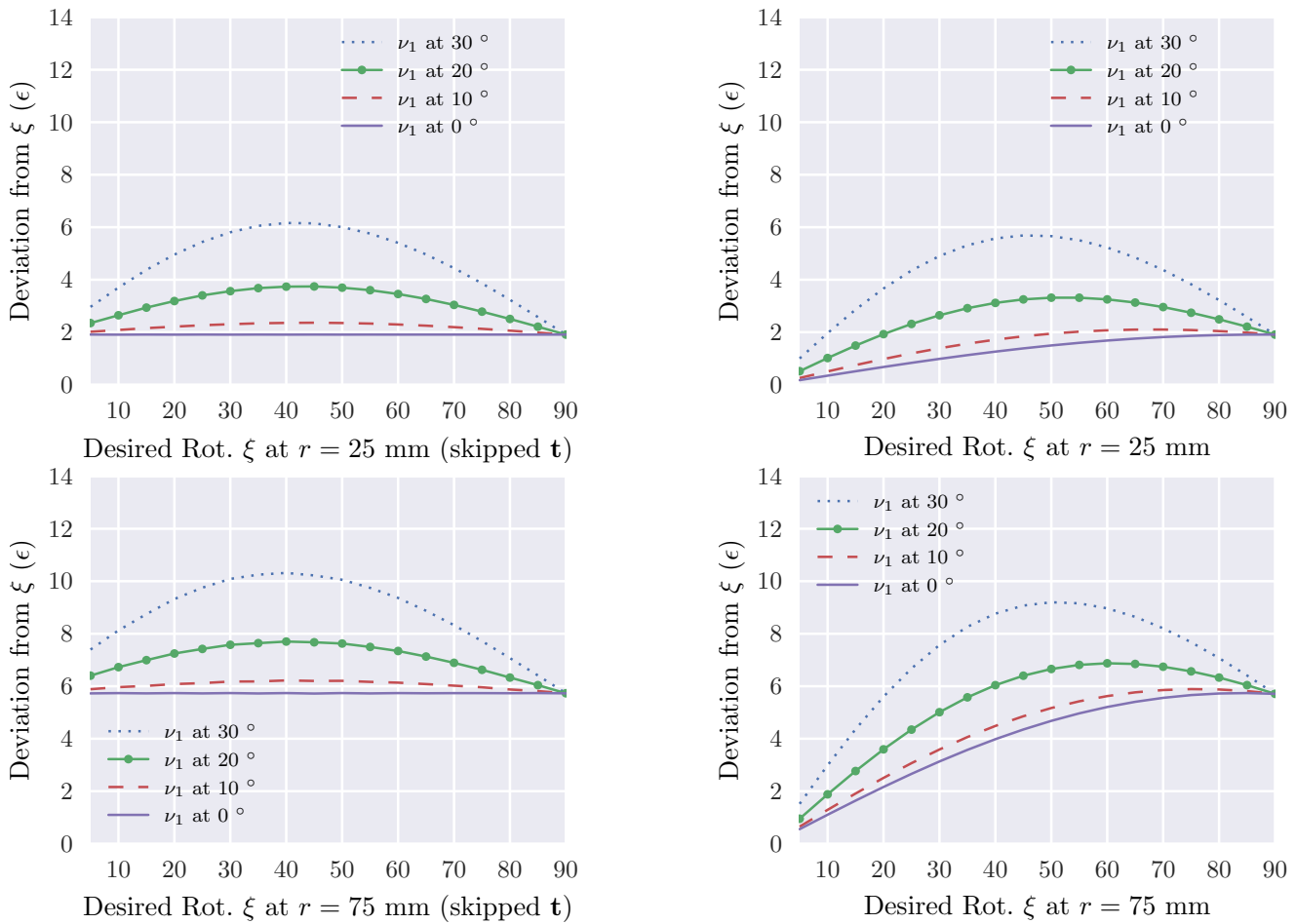

	\centering %45
	\resizebox{0.45\linewidth}{!}{
		\input{figure4a}}
	\hfill \centering
	\resizebox{0.45\linewidth}{!}{
		\input{figure4b}}
	\hfill \centering
	\resizebox{0.45\linewidth}{!}{
		\input{figure4c}}
	\hfill \centering
	\resizebox{0.45\linewidth}{!}{
		\input{figure4d}}
	\caption{
		Plots depicting $\epsilon$ as a function $\xi$. The two plots in the left column correspond to the algorithm with skipped translation, and the two plots in the right column correspond to the full algorithm.}
	\label{fig:xi_accuracy}
\end{figure*}
\label{evaluation}
Our viewpoint planning algorithm uses a single \mbox{2-D} angiographic image only to optimally integrate with the current clinical workflow and we evaluate the method accordingly. We identify the reliability of calculated viewpoints based on different inputs to the method, and state results for both the complete and the rotation-only method. 
In addition, we implemented the proposed algorithm on a C-arm system. Using a phantom of the left coronary artery tree (LCA) we investigate the impact of input views to the quality of QCA. 
\subsection{Accuracy of 2-D Viewpoint Planning}

The first important quantity on view selection for providing high quality QCA  is the true rotation around the vessel $\phi = \xi + \epsilon$ with the desired rotation $\xi$ and some unwanted rotation $\pm\epsilon$. Typically an additional rotation might not always harm the result but especially when a minimum angulation must be fulfilled, a high precision is important.  The second quantity is the foreshortening of the vessel within each of the two views $\nu_1$ and $\nu_2$. 
The physician is responsible for $\nu_1$, whereas our algorithms can only influence $\nu_2$.
In our evaluation we inspect the angular precision measured in $\epsilon$ and the foreshortening  $\nu_2$ based on given desired rotations $\xi$  and different positions of the vessel segment in the initial frame that determines $\nu_1$. 

Covering all possible line configurations seems impractical, and we decide to parameterize the vessel segment position using an offset from the origin $r$ and foreshortening $\nu_1$. Based on such a line configuration we perform the proposed algorithm. Since the true scene is known we can then calculate $\epsilon$ and $\nu_2$. However, a line is not uniquely defined by $\nu_1$ and an offset $r$ as the offset could e.g. be parallel or orthogonal to the detector.  Therefore, the stated $\epsilon$ and $\nu_2$ represent the worst case foreshortening and angulation inaccuracy over all lines having a certain offset $r$ and foreshortening $\nu_1$.% 

\subsubsection{Foreshortening}
\begin{figure*}
	\centering     %%% not \center
	\subfigure[RAO: 27, CAUD: 11.]
	{
		\label{subfig:default}
		\includegraphics[width=.3\textwidth]{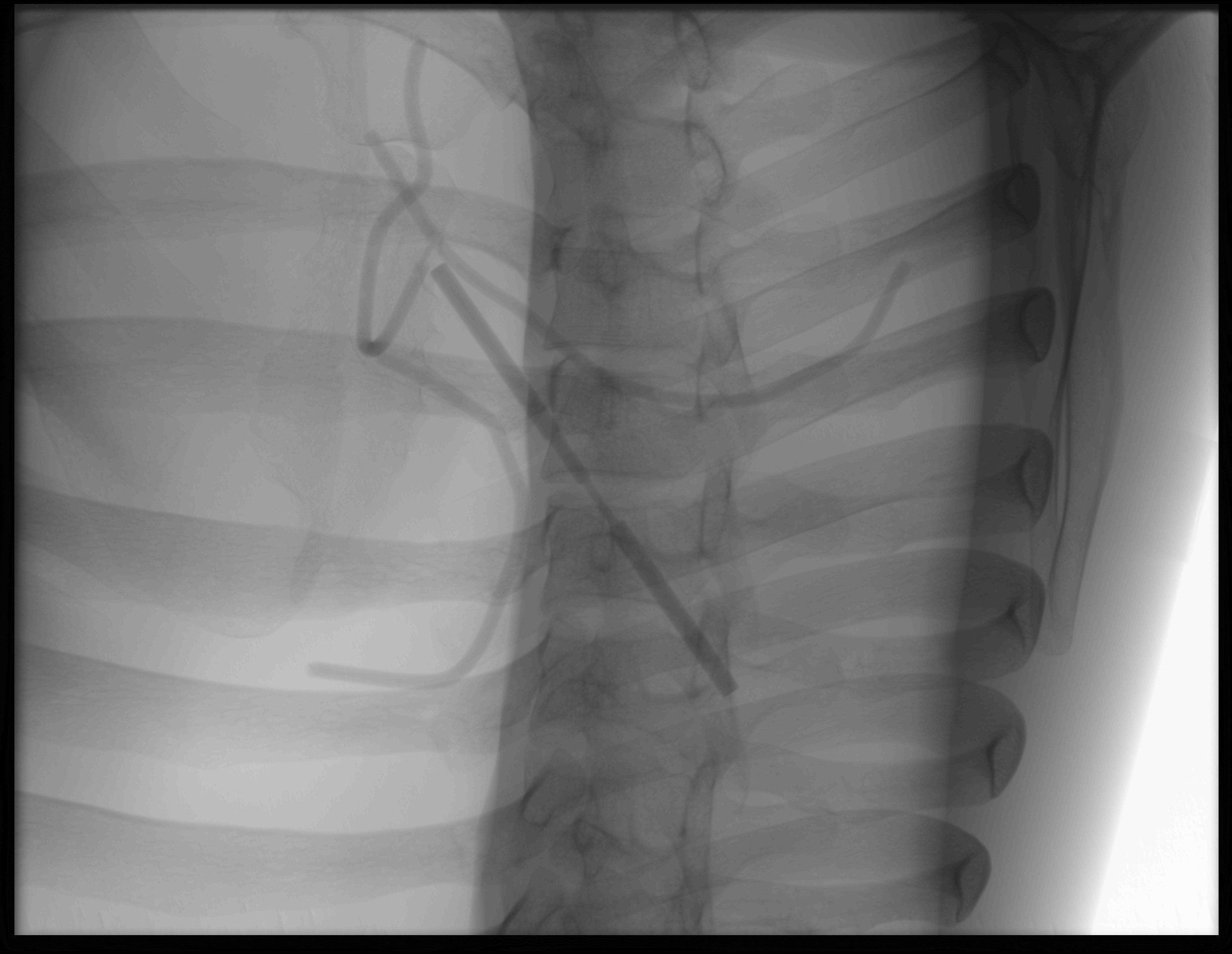}
	}
	\subfigure[LAO: 6, CAUD: 30.]
	{
		\label{subfig:vpp}
		\includegraphics[width=.3\textwidth]{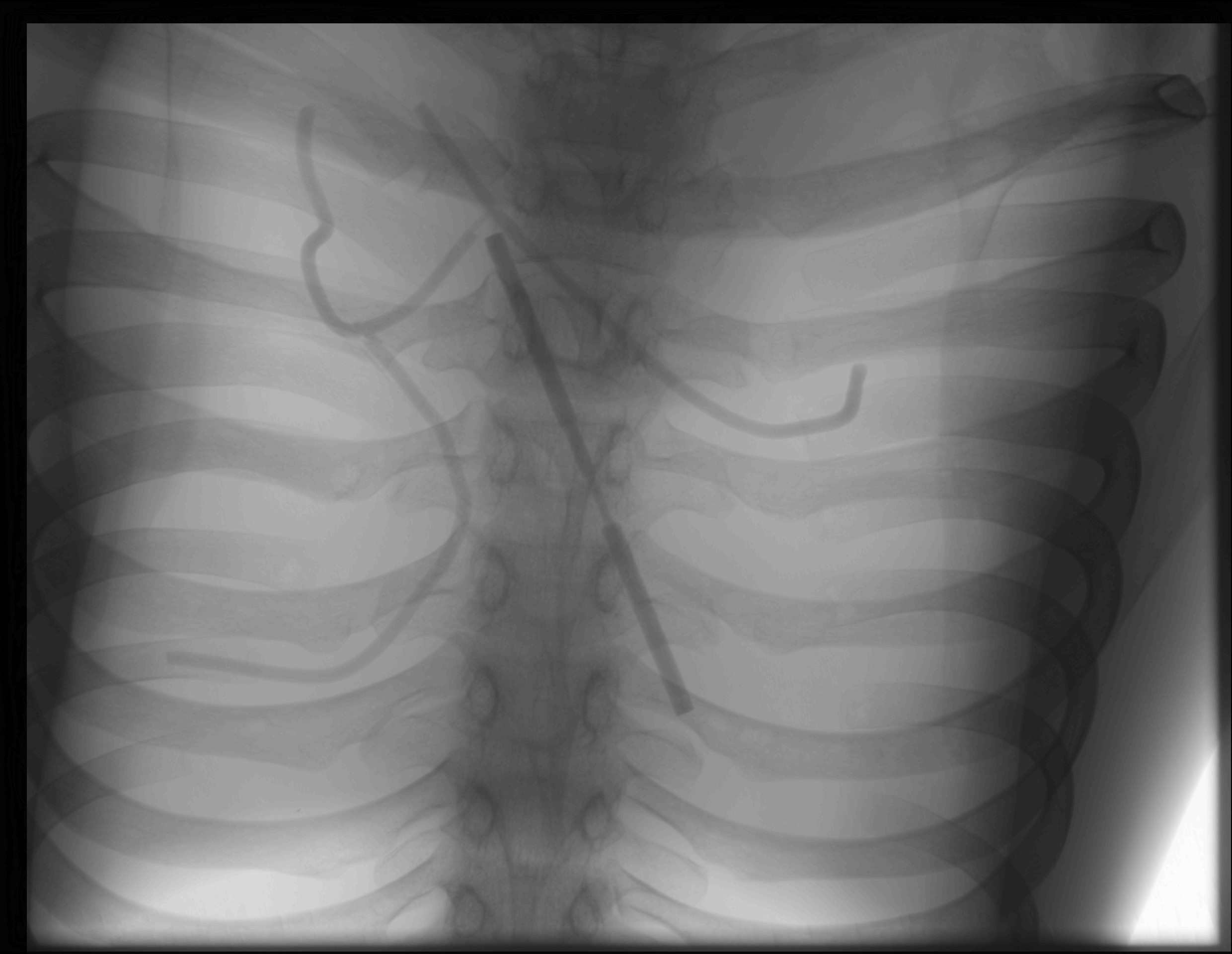}
	}
	\subfigure[RAO: 15, CRAN: 13.]
	{
		\label{subfig:manual}
		\includegraphics[width=.3\textwidth]{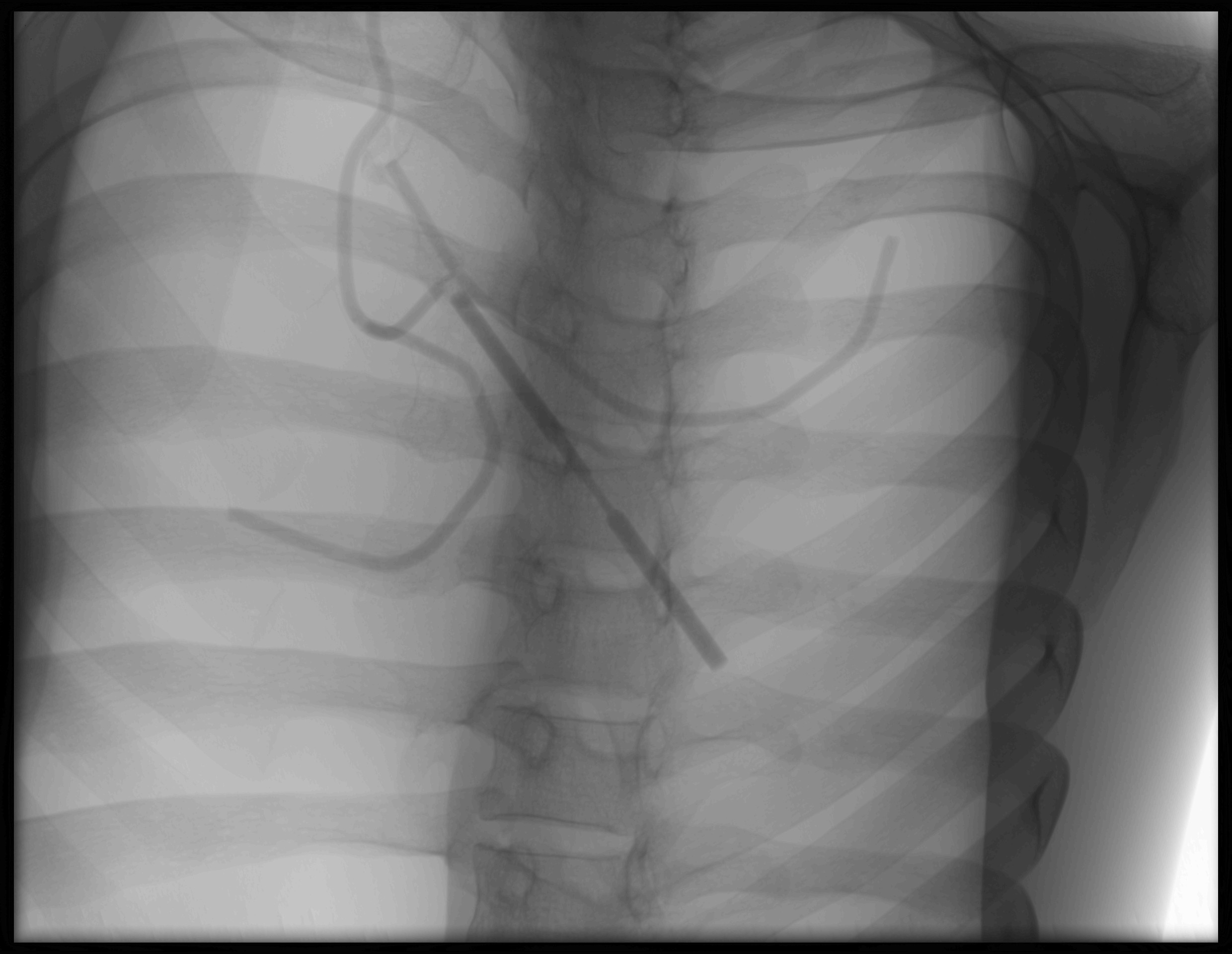}
	}
	\subfigure[QCA using \ref{subfig:default} and \ref{subfig:vpp} as inputs.]
	{
		\label{subfig:area_good}
		\includegraphics[width=.45\textwidth]{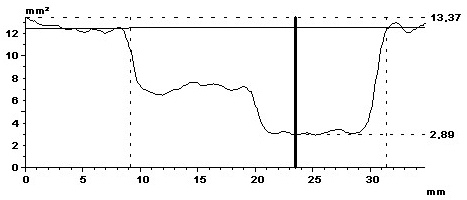}
		
	}
	\subfigure[QCA using \ref{subfig:default} and \ref{subfig:manual} as inputs.]
	{
		\label{subfig:area_bad}
		\includegraphics[width=.45\textwidth]{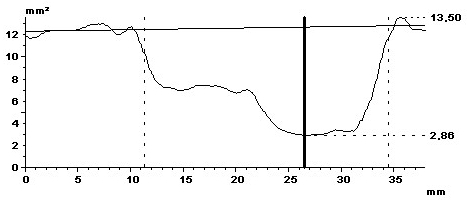}
	}
	\caption{Top row: Input views for QCA generation. The initial view (a), a second view generated using the proposed viewpoint planning with $\xi = 35\degree$ (b), and a manual selected second view (c). Bottom: Area curves deduced from a 3-D QCA using two input views. The curves are exported from the clinical report generated with the 3-D QCA tool \textit{syngo} IZ3D (Siemens Healthcare GmbH, Forchheim).}
	\label{fig:phantom}
\end{figure*}

The impact of different line configurations on the foreshortening in the second view is depicted in Fig. \ref{fig:fs_accuracy}. 
A very practical aspect is the elimination of foreshortening when requesting a 90$\degree$ rotation. This elimination is independent on the positioning of the vessel and can therefore be guaranteed. In general, low $\nu_2$ values are associated with high $\xi$ values, suggesting that large angulations are desirable if a substantial reduction in foreshortening needs to be achieved. For all clinically relevant cases where the vessel off-center displacement is very likely below $100\,$mm, our algorithm consistently reduces foreshortening. The influence of an offset $r$ to $\nu_2$ is negligible small.

\subsubsection{Angular Precision}

The angular precision is depicted in Fig. \ref{fig:xi_accuracy}. It is observable that a high foreshortening $\nu_1$ reduces the angular precision, whereas the most inaccuracy occurs at $\xi\approx40\degree$. The translation is capable of increasing the angular accuracy at small $\xi$ values; however, for large $\xi$ values the translation only minimally affects the result.  

The two plots in the upper row are created with a fixed $r = 25\,$mm, whereas the lower two plots are created with a fixed $r = 75\,$mm. 
This off-centering is mostly visible by an overall offset of angular precision. 

\subsection{Phantom Study}
To inspect the impact of input views to the quality of \mbox{3-D} QCA, we placed an aluminum phantom of the LCA in a thorax phantom. The vessel segment of interest is a commercially available calibration phantom consisting of three piecewise constant diameters that are successively narrowing to mimic a stenosis. The quality is inspected using the area curves of the reconstructed vessel (cf. Fig. \ref{fig:phantom}). 

We generate two QCAs of the phantom both generated from two views, one using the proposed algorithm and the other using manual selected views, whereas both use a similar amount of rotation. The input view in Fig. \ref{subfig:default} is used as the initial view for both QCAs. The view depicted in Fig. \ref{subfig:vpp} is created using an implementation of the viewpoint planning algorithm on a C-arm system with $\xi = 35\degree$. The resulting area curve is depicted in Fig. \ref{subfig:area_good}. The area curve in Fig. \ref{subfig:area_bad} is generated using the manually selected view depicted in Fig. \ref{subfig:manual}.

As the wire phantom is piecewise constant and radial symmetric, it is not to be expected that constant regions are strongly view dependent, however, in the transitions to narrower or brighter diameters a view dependence is observable. Using the viewpoint planning the transitions (cf. Fig. \ref{subfig:area_good} at 9.1, 20 and 31.3~mm) are much sharper and well defined, whereas the manually selected views produce a smearing of the transitions (cf. Fig. \ref{subfig:area_good} at 11.2, 23 and 33~mm).

\section{Outlook and Discussion}
\label{discussion}
Our accuracy evaluation states the worst possible outcome when our planning algorithm is used. The largest errors typically occur if the vessel is translated in the viewing direction as a displacement  that cannot easily be recovered from a single projection image. 
Rotations around vessels that are translated parallel to the detector are achieved with much higher precision and, in the best case, even exact. However, using the worst case experiments, we can ensure certain minimum requirements.
For instance, consider a required minimum rotation around the segment of $\xi = 30\degree$, e.g. to ensure \mbox{3-D} reconstruction of acceptable quality: if the physician can ensure that the target vessel is foreshortened by less then $20\degree$ and that it is located within a  $25\,$mm radius around the isocenter, then we can accurately calculate the required transformation. An angulation of $33\degree$ is sufficient to ensure a $30\degree$ rotation. When translation is not possible, the $30\degree$ angulation is ensured with a $34\degree$ rotation. 
These minimum requirements are often important in clinical practice, as the C-arm can be restricted in its movements either due to the patient or the anatomy, making a simple $90\degree$ rotation infeasible.

Our phantom study showed that changes in the diameter are smeared if the view selection is not taken carefully. This comes in favor to the results presented by Sato et al. \cite{Sato1998}. Particularly, in a real clinical setup these inaccuracies can limit the reliability of QCA especially for small stenosis.

A problem not yet addressed by the algorithm is the overlapping of vessels as well as vessels leaving the field-of-view. Possible improvements for overcoming these drawbacks could be the  extension of the method  with prior knowledge, e.g. by favoring viewpoints that are empirically known to produce good results (see for example \cite{Garcia2009,Wink2002}).

To conclude, we introduced a method for viewpoint planning in coronary angiography based on a single \mbox{2-D} image. If two or more images have been acquired, making a \mbox{3-D} centerline reconstruction possible, exact methods~---~e.g. \cite{Chrisriaens2001,Garcia2009,JamesChen2000}~---~will outperform the proposed planning system. The proposed method is therefore of advantage if no 3-D information is available, or cannot be utilized due to a missing system-patient registration. Our algorithms allow personalized planning of standardized views in conventional angiography that could translate to reduced dose to the patient and operating team while promoting improved quality for QCA. 

~\\
\textbf{Disclaimer} The concepts and information presented in this paper are based
on research and are not commercially available.

~\\
\textbf{Conflict of interest} S. Achenbach, M. Unberath and A. Maier have no conflict of interest. A. Preuhs is funded by Siemens Healthcare GmbH, Forchheim Germany. M. Berger, S. Bauer and T. Redel are employees of Siemens  Healthcare GmbH, Forchheim Germany. 

~\\
\textbf{Informed consent} This article does not contain patient data.


\begin{thebibliography}{10}
	
	\bibitem{Blinn1977}
	J.~Blinn.
	\newblock {A Homogeneous Formulation for Lines in 3 Space}.
	\newblock In {\em Siggraph 1977}, pages 237--241. Association for Computing
	Machinery, Inc., 1977.
	
	\bibitem{Chrisriaens2001}
	J.~Chrisriaens, R.~Walle, and I.~Lemahieu.
	\newblock {A Simple Determination System For Optimal Angiographic Viewing}.
	\newblock In {\em Conf. on Image Processing}, volume~2, pages 327--330, 2001.
	
	\bibitem{delaere1991}
	D.~Delaere, C.~Smets, P.~Suetens, G.~Marchal, and F.~Van~de Werf.
	\newblock Knowledge-based system for the three-dimensional reconstruction of
	blood vessels from two angiographic projections.
	\newblock {\em MBEC},
	29(6):NS27--NS36, 1991.
	
	\bibitem{fallavollita2014}
	P.~Fallavollita, A.~Winkler, S.~Habert, P.~Wucherer, P.~Stefan, R.~Mansour,
	R.~Ghotbi, and N.~Navab.
	\newblock Desired-view controlled positioning of angiographic c-arms.
	\newblock In {\em MICCAI}, pages 659--666. Springer, 2014.
	
	
	\bibitem{Garcia2009}
	J.~Garcia, B.~Movassaghi, I.~Casserly, A.~Klein, S.~Chen,
	J.~Messenger, A.~Hansgen, O.~Wink, B.~Groves, and J.~Carroll.
	\newblock {Determination of optimal viewing regions for X-ray coronary
		angiography based on a quantitative analysis of 3D reconstructed models}.
	\newblock {\em Cardiovascular Imaging}, 25(5):455--462, 2009.
	
	\bibitem{Green2005}
	N.~Green, S.~Chen, A.~Hansgen, J.~Messenger, B.~Groves, and
	J.~Carroll.
	\newblock {Angiographic views used for percutaneous coronary interventions: A
		three-dimensional analysis of physician-determined vs. computer-generated
		views}.
	\newblock {\em Catheterization and Cardiovascular Interventions},
	64(4):451--459, 2005.
	
	\bibitem{JamesChen2000}
	S.~{Chen} and J.~Carroll.
	\newblock {3-D reconstruction of coronary arterial tree to optimize
		angiographic visualization}.
	\newblock {\em Tansactions on Medical Imaging}, 19(4):318--336, 2000.
	
	\bibitem{Liu1992}
	I.~Liu and Y.~Sun.
	\newblock Fully automated reconstruction of three-dimensional vascular tree
	structures from two orthogonal views using computational algorithms and
	productionrules.
	\newblock {\em Optical Engineering}, 31(10):2197--2208, 1992.
	
	\bibitem{Mozaffarian2015}
	D.~Mozaffarian, E.~Benjamin, A.~Go, D.~Arnett, M. Blaha, M.~Cushman, ... and V.~Howard
	\newblock {Executive Summary: Heart Disease and Stroke Statistics 2015 Update}.
	\newblock {\em Circulation}, 131(4):434--441, 2015.
	
	\bibitem{Pellot1994}
	C.~Pellot, M.~Sigelle, P.~Horain, and P.~Peronneau.
	\newblock {A 3D Reconstruction of Vascular Structures from Two X-Ray Angiograms
		Using an Adapted Simulated Annealing Algorithm}.
	\newblock {\em TMI}, 13(1):48--60, 1994.
	
	\bibitem{Sato1998}
	Y.~Sato, T.~Araki, M.~Hanayama, H.~Naito, and S.~Tamura.
	\newblock {A viewpoint determination system for stenosis diagnosis and
		quantification in coronary angiographic image acquisition.}
	\newblock {\em TMI}, 17(1):121--37, 1998.
	
	\bibitem{Tu2010}
	S.~Tu, G.~Koning, W.~Jukema, and J.~H. Reiber.
	\newblock Assessment of obstruction length and optimal viewing angle from
	biplane x-ray angiograms.
	\newblock {\em The int. journal of cardiovascular imaging},
	26(1):5--17, 2010.
	
	
	\bibitem{Unberath2017}
	M.~Unberath, O.~Taubmann, M.~Hell, S.~Achenbach, and A.~Maier.
	\newblock {Symmetry, Outliers, and Geodesics in Coronary Artery Centerline
		Reconstruction from Rotational Angiography}.
	\newblock {\em Medical Physics}, 44(11):5672--85, 2017.
	
	\bibitem{virga2015}
	S.~Virga, V.~Dogeanu, P.~Fallavollita, R.~Ghotbi, N.~Navab, and S.~Demirci.
	\newblock Optimal c-arm positioning for aortic interventions.
	\newblock In {\em BVM 2015}, pages 53--58.
	Springer, 2015.
	
	\bibitem{Wink2002}
	O.~Wink, R.~Kemkers, S.~Chen, and J.~D. Carroll.
	\newblock Coronary intervention planning using hybrid 3d reconstruction.
	\newblock In {\em MICCAI}, pages 604--611. Springer, 2002.
	
\end{thebibliography}
\end{document}